\documentclass[fleqn,usenatbib]{mnras}

\usepackage{newtxtext,newtxmath}

\usepackage[T1]{fontenc}

\DeclareRobustCommand{\VAN}[3]{#2}
\let\VANthebibliography\thebibliography
\def\thebibliography{\DeclareRobustCommand{\VAN}[3]{##3}\VANthebibliography}

\usepackage{graphicx}	
\usepackage{amsmath}	

\title[The Velocity Statistics of Turbulent Clouds]{The Velocity Statistics of Turbulent Clouds in the Presence of Gravity, Magnetic fields, Radiation, and Outflow Feedback}

\author[Hu et al.]{
Yue Hu$^{1,2}$\thanks{E-mail: yue.hu@wisc.edu}
,Christoph Federrath$^{3}$\thanks{E-mail: christoph.federrath@anu.edu.au}
,Siyao Xu$^{4}$\thanks{E-mail: sxu@ias.edu (Hubble Fellow)}
,Sajay Sunny Mathew$^{3}$
\\
$^{1}$Department of Physics, University of Wisconsin-Madison, Madison, WI, 53706, USA\\
$^{2}$Department of Astronomy, University of Wisconsin-Madison, Madison, WI, 53706, USA\\
$^{3}$Research School of Astronomy and Astrophysics, Australian National University, Canberra, ACT 2611, Australia\\
$^{4}$Institute for Advanced Study, 1 Einstein Drive, Princeton, NJ 08540, USA\\
}

\date{Accepted XXX. Received YYY; in original form ZZZ}

\pubyear{2022}

\begin{document}
\label{firstpage}
\pagerange{\pageref{firstpage}--\pageref{lastpage}}
\maketitle
\begin{abstract}
The interaction of turbulence, magnetic fields, self-gravity, and stellar feedback within molecular clouds is crucial for understanding star formation. We study the effects of self-gravity and outflow feedback on the properties of the turbulent velocity via the structure function over length scales from $\sim$ 0.01 pc to 2 pc. We analyze a series of three-dimensional, magnetohydrodynamical (MHD) simulations of star cluster formation. We find outflow feedback can change the scaling of velocity fluctuations but still roughly being in between Kolmogorov and Burgers turbulence. We observe that self-gravity and protostellar outflows increase the velocity fluctuations over all length scales. Outflows can amplify the velocity fluctuations by up to a factor of $\sim$7 on scales $\sim$ 0.01 - 0.2 pc and drive turbulence up to a scale of $\sim$ 1 pc. The amplified velocity fluctuations provide more support against gravity and enhance fragmentation on small scales. The self-gravity's effect is more significant on smaller dense clumps and it increases the fraction of the compressive velocity component up to a scale of $\sim$ 0.2 pc. However, outflow feedback drives both solenoidal and compressive modes, but it induces a higher fraction of solenoidal modes relative to compressive modes. Thus, with outflows, the dense core ends up with a slightly higher fraction of solenoidal modes. We find that the compressible fraction is fairly constant with about 1/3 on scales $\sim$ 0.1 - 0.2 pc. The combined effect of enhanced velocity dispersion and reduced compressive fraction contributes to a reduction in the star formation rate. 
\end{abstract}




\section{Introduction}
Understanding how stars form requires knowledge of the interplay between turbulence, magnetic fields, self-gravity, and outflow feedback within molecular clouds and star-forming sub-regions. In the multi-phase interstellar medium (ISM; \citealt{2001RvMP...73.1031F,2020PPCF...62a4014F}), turbulence permeates over a extensive range of length scales being either nearly incompressible or highly compressive \citep{1995ApJ...443..209A,2004ApJ...615L..45H,2010A&A...512A..81F,2010ApJ...710..853C,2017ApJ...835....2X,2019NatAs...3..154L,2020ApJ...901..162H,2021NatAs...5..365F}. In particular, turbulence can both provide global support against gravitational collapse on the scales of molecular clouds \citep{1993ApJ...419L..29E,1995MNRAS.277..377P,2000ApJ...535..887K} and produce local density fluctuations serving as seeds of star formation \citep{MacLowKlessen2004,2005ApJ...630..250K,2007ARA&A..45..565M}. Magnetic fields, however, uni-directionally work against the cloud being gravitationally contracted \citep{2012ApJ...761..156F,2020ApJ...890..157X,2020ApJ...897..123H}.

Moreover, self-gravity can alter turbulence and magnetic field statistics. For instance, gravitational collapse produces a shallower density power spectrum \citep{KimRyu2005,FederrathKlessen2013,2015ApJ...808...48B,2018ApJ...856..136P,2019A&A...630A..97C,2020ApJ...905..129H} and drives small-scale turbulence \citep{NakamuraLi2007,KlessenHennebelle2011,FederrathEtAl2011,KrumholzBurkhert2016}. It may also shape a weak magnetic field to be hourglass-like \citep{2013ApJ...767...33E,2019ApJ...885..106L}. Moreover, stellar radiation and outflow feedback may drive supersonic turbulence \citep{2009ApJ...695.1376C}, reduce the star formation rate by about a factor of $\sim2$, and shift the initial mass function (IMF) to lower masses by a factor of $\sim3$ \citep{2014ApJ...790..128F,2021MNRAS.507.2448M,GrudicEtAl2021}. However, the effects of self-gravity and outflows on the statistical properties of turbulence have not been studied in much detail. As density fluctuations alone fail to distinguish between fossil and active turbulence, velocity fluctuations are more direct and dynamically important measurements of turbulence \citep{2003MNRAS.342..325E}, which are the focus of this work.

The structure function (SF) is one of the most commonly used statistical tool to access the properties of velocity fluctuations. For nearly incompressible turbulence, such as in the diffuse ISM, velocity fluctuations follow Kolmogorov's $v_\ell \propto \ell^{1/3}$ scaling, where $\ell$ denotes the scale of interest. Consequently the second-order SF appears as a power law $\sim\ell^{2/3}$ \citep{2003MNRAS.345..325C,2018ApJ...864..116Q,2021ApJ...911...37H,2021ApJ...910...88X,2021arXiv211115066H}. The power law, however, may change to $\sim \ell^1$ for supersonic, highly-compressible turbulence \citep{KritsukEtAl2007,SchmidtEtAl2009,FederrathEtAl2009,2010A&A...512A..81F,2010ApJ...720..742K,KonstandinEtAl2012,2021NatAs...5..365F,2021arXiv211115066H}. In addition, in a self-gravitating medium, infall motions induced by gravitational collapse and outflow feedback from star formation can introduce extra velocity fluctuations. These effects may leave an imprint on the SF by changing its amplitude and/or slope. Therefore, one may be able to use the SF to obtain information about the structure and/or the evolution of star-forming regions. In particular, stellar feedback, such as stellar winds and supernova explosions, inject turbulent kinetic energy and replenish the turbulence \citep{NakamuraLi2007,2011ApJ...738...88H,2014ApJ...790..128F,2015MNRAS.450.4035F}. The turbulence amplified by outflow feedback could provide extra support against gravity and enhance fragmentation on small scales.

This work aims at determining the role of gravity and stellar outflow feedback for turbulence. We use a series of three-dimensional, magnetohydrodynamical (MHD) simulations of star cluster formation including self-gravity, turbulence, magnetic fields, stellar radiative heating, and outflow feedback. These simulations were developed by \cite{2021MNRAS.507.2448M}.

The paper is organized as follows. In \S~\ref{sec:method}, we provide the details of the simulation data and illustrate the SF method used in this work. In \S~\ref{sec:results}, we study the effect of jets/outflows in regulating the SF by comparing simulations that include outflow feedback with the same simulations, but without protostellar outflows. We also investigate the evolution of the compressive velocity field in the presence of outflow feedback. We discuss  summarize the results in \S~\ref{sec:con}, respectively.

\section{Methods}
\label{sec:method}
\subsection{Numerical simulations}

The numerical simulations were developed by \citet{2021MNRAS.507.2448M}. We briefly describe the numerical methodology here. The cloud is modelled by the FLASH code \citep{2000ApJS..131..273F,2008ASPC..385..145D}. FLASH solves the magnetohydrodynamical (MHD) equations including gravity on an adaptive mesh refinement (AMR; \citealt{1989JCoPh..82...64B}) grid using the PARAMESH library \citep{2000CoPhC.126..330M}. For these simulations, we use the HLL3R MHD solver \citep{WaaganEtAl2011}.

The cloud is simulated in a three-dimensional triple-periodic box with side length $L=2$~pc. The maximum refinement level provides a maximum effective grid resolution of $4096^3$ cells or minimum cell size of 100~AU. AMR is triggered when the local (cell-by-cell) Jeans length drops below 16~grid cell lengths on any level of refinement other than the maximum level \citep{FederrathEtAl2011}. On the maximum level of AMR, we introduce sink particles \citep{2010A&A...512A..81F} to model local collapse and accretion of star-forming cores. The initial gas density is uniformly set to $\rho_0=6.56 \times 10^{-21}$~g~$\mathrm{cm}^{-3}$, which gives a total cloud mass of 775~$M_\odot$ and a mean free-fall time of $t_{\rm ff} =0.82$~Myr. The initial magnetic field $B=10^{-5}$~G is uniform along the $z$-axis.

The turbulent acceleration field continuously drives turbulent motions. It is modelled by a stochastic Ornstein-Uhlenbeck process \citep{1988CF.....16..257E,2010A&A...512A..81F}. Initially, in the absence of self-gravity, kinetic energy is injected on scales that correspond to wavenumbers $k=[1,3]$, where $k$ is measured in units of $2\pi/L$, such that the driving amplitude is largest at $k=2$ and drops to zero on either side of $k=2$, following a parabolic spectrum as used in previous studies \citep{FederrathKlessenSchmidt2008,SchmidtEtAl2009,2010A&A...512A..81F,Federrath2013,2021NatAs...5..365F}. In particular, mixed turbulence driving is implemented. It naturally results in about 1/3 fraction of compressive power and a power-law slope of $\sim-2$ for the velocity power spectrum \citep{KritsukEtAl2007,2010A&A...512A..81F,Federrath2013,2021NatAs...5..365F}, in the regime of supersonic turbulence, typically found to be appropriate for the dense, cold phase of the ISM \citep{Larson1981,SolomonEtAl1987,OssenkopfMacLow2002,2004ApJ...615L..45H,RomanDuvalEtAl2011}. The initial sonic Mach number of 5 is set by the velocity dispersion of $1.0$ km s$^{-1}$ and the initially isothermal sound speed of $c_s=0.2$~km~s$^{-1}$.

After turbulence is fully developed ($\sim2$~Myr), self-gravity is switched on. Here gravity in the equations is contributed by gas and sink particles. The self-gravity of the gas is calculated via a multi-grid Poisson solver \citep{2008ApJS..176..293R}. Also, a polytropic equation of state $P=c_s^2\rho^\gamma$ for the gas pressure is used, where $c_s$ is the sound speed and $\rho$ is the gas density. The value of the polytropic exponent $\gamma$ varies with the local density of the gas and is based on previous detailed radiation-hydrodynamic simulations of protostar formation (see \citealt{2021MNRAS.507.2448M} for details). The polar stellar heating model developed by \cite{2020MNRAS.496.5201M} is implemented to include protostellar heating in the simulations.
 
Sink particles are introduced in regions that are undergoing gravitational collapse, as verified by an automatic procedure involving a number of checks performed in a control volume of radius $r_{\rm sink}$ around anby computational cell that exceeds the threshold density defined by the local Jeans length,
\begin{equation}
    \rho_{\rm sink}=\frac{\pi c_s^2}{G\lambda_J^2}=\frac{\pi c_s^2}{4Gr_{\rm sink}^2},
\end{equation}
where $G$ is the gravitational constant, $\lambda_J=\sqrt{\pi c_s^2/(G\rho)}$ is the local Jeans length, and $r_{\rm sink}=\lambda_J/2$ is the sink particle radius. The size of sink particles is defined to ensure that the Truelove criterion \citep{1997ApJ...489L.179T} is satisfied on the highest level of AMR, i.e., $2r_{\rm sink}=5\Delta x$, where $5\Delta x$ is the grid cell length on the highest level of refinement.

Finally, use the subgrid-scale (SGS) jet/outflow model implemented in \cite{2014ApJ...790..128F} to launch jets and outflows from sink particles. The outflow model produces a fast collimated jet component, and a lower-speed, wider outflow component as typically observed for protostellar jets/outflows. It transfers mass, momentum, and angular momentum back into the parental cloud, with parameters that were physically calibrated via dedicated high-resolution jet simulations, theoretical models of jets launching, and observational data \citep{2014ApJ...790..128F}.

\begin{table*}
	\centering
	\label{tab:sim}
	\setlength{\tabcolsep}{3pt}
	\begin{tabular}{| c | c | c | c | c | c | c | c | c| c| }
		\hline
		Model & Jets/Ouflows & $t_{5\%}$ [$t_{\rm ff}$] & $\overline{\rm SFR}_{\rm ff}\;[\%]$ & $N_{\rm sinks}$&  $\langle\sigma_v^{\rm 5\%}\rangle_{0.01-0.1}$ [km/s] &  $\langle\sigma_v^{5\%}\rangle_{0.1-0.2}$ [km/s] & $\langle\chi^{\rm 5\%}\rangle_{0.01-0.1}$ & $\langle\chi^{\rm 5\%}\rangle_{0.1-0.2}$\\	\hline\hline
		NOWIND & No & $0.68\pm0.15$ & $15\pm3$ & 212 & $0.61\pm0.16$& $0.93\pm0.22$ & $0.30\pm0.01$ & $0.34\pm0.04$\\
		OUTFLOW & Yes & $0.89\pm0.2$ & $7\pm2$ & 449 & $1.99\pm1.17$ & $2.62\pm1.68$ & $0.29\pm0.01$ & $0.32\pm0.02$\\
		\hline
	\end{tabular}
	\caption{Key simulation parameters and results. Ten simulations with different turbulence realisations (T1--T10) are run for both the NOWIND and OUTFLOW models. In the table, $t_{5\%}$ is the average time taken (in units of the free-fall time) by the simulations to reach SFE = 5\% and is measured from the moment self-gravity is turned on. The value of $\overline{\rm SFR}_{\rm ff}$ quoted in the table is time average. 
	$\langle\sigma_v^{\rm tur.}\rangle$ (and $\langle\chi^{\rm tur.}\rangle$) and $\langle\sigma_v^{5\%}\rangle$ (and $\langle\chi^{\rm 5\%.}\rangle$) are calculated (see Figs.~\ref{fig:frac_long04} and \ref{fig:Efrac04}) for the snapshots "fully developed turbulence" and "SFE = 5\%", respectively.	
	The subscripts $"0.01-0.1"$ and $"0.1-0.2"$ means the values are averaged over $0.01-0.1$~pc and $0.1-0.2$~pc, respectively. For both model, $\langle\sigma_v^{\rm tur.}\rangle_{0.01-0.1}=0.30\pm0.09$, $\langle\sigma_v^{\rm tur.}\rangle_{0.1-0.2}=0.51\pm0.17$, $\langle\chi^{\rm 5\%}\rangle_{0.01-0.1}=0.27\pm0.01$, and $\langle\chi^{\rm 5\%}\rangle_{0.1-0.2}=0.29\pm0.03$.
	The resolution level, cloud properties and turbulence setup are the same in both models and the only difference is that protostellar jets/outflows are absent in the NOWIND simulations.}
\end{table*}
\begin{figure*}
	\centering
	\includegraphics[width=1.0\linewidth]{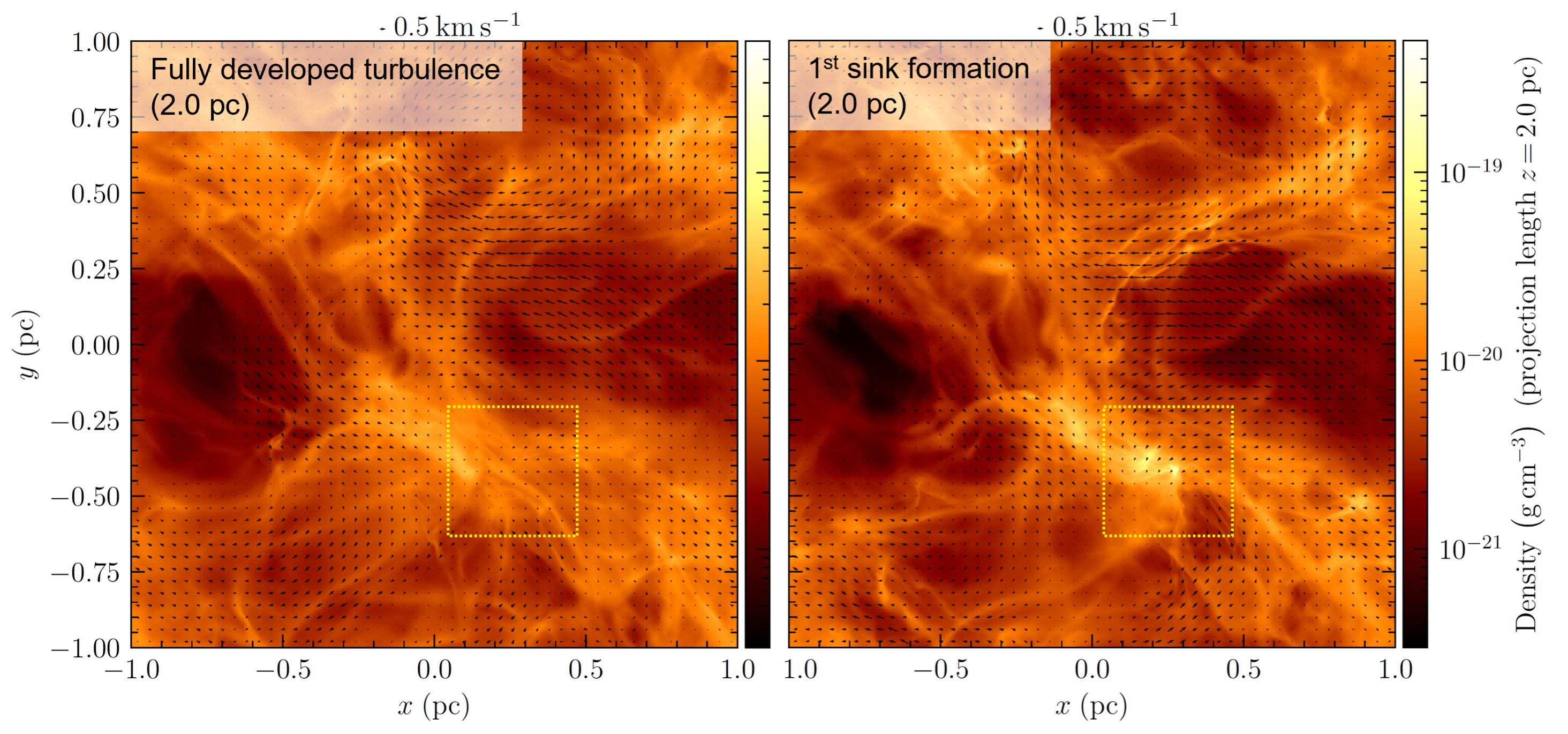}
	\caption{\label{fig:illu} Gas column density in one of the ten simulations (the T2 simulation, i.e., with turbulence seed \#2), at the stage of fully developed turbulence (left-hand panel) and just before the $1^{\rm st}$ sink particle has formed (right-hand panel). The black vectors represent the projected velocity field and their length indicates the amplitude of the local velocity. The yellow dashed rectangle outlines the zoom-in region used for structure function calculations below.}
\end{figure*}

Here we primarily compare two simulation models: the NOWIND and the OUTFLOW models from \citet{2021MNRAS.507.2448M}. Both of them have the same initial conditions. However, the OUTFLOW simulation set additionally includes jet and outflow feedback from the formed stars. As in \citet{2021MNRAS.507.2448M} we use a total of ten simulations with different turbulence realizations (T1--T10) for both the NOWIND and OUTFLOW models, which allows us to draw statistically significant results and provides us with a quantification of the variations between different turbulent realizations. Tab.~\ref{tab:sim} lists the main simulation parameters and derived results in this work, as discussed in detail below.

\subsection{The second-order structure function}
In this work, we aim to quantify the turbulent velocity statistics of star-forming regions, with a particular focus on the role of outflows feedback. In particular, we measure the structure function, which is a commonly used method to quantify the statistical properties of turbulent flows. The second-order structure function is defined as  
\begin{equation}
    {\rm SF_2}(r)=\langle|\pmb{v}(\pmb{x}+\pmb{r})-\pmb{v}(\pmb{x})|^2\rangle_r,
\end{equation}
where $\pmb{v}(\pmb{x})$ represents the velocity at the spatial position $\pmb{x}$ and $\pmb{r}$ is the separation vector \footnote{In fact, the ${\rm SF_2}(r)$ as a function of $\pmb{r} $ can trace magnetic field orientation and strenght\citep{2021ApJ...910...88X,2021ApJ...911...37H,2021ApJ...915...67H}.}. The structure function ${\rm SF_2}(r)$ takes the ensemble average $\langle...\rangle_r$ over a sufficiently large sample at the same separation value $r$. The structure function was computed and averaged
over the set of 10~simulations that are identical in all physical parameters, but used different random seeds for the turbulence. Its uncertainty is given by the standard deviation. In particular, we take the square root of the normalized structure function, i.e.,
\begin{equation}
\sigma_v(r) = \sqrt{{\rm SF}_2(r)/2},
\label{eq:sfnorm}
\end{equation}
which represents the 3D velocity dispersion on scale $r$.
\begin{figure*}
	\centering
	\includegraphics[width=0.99\linewidth]{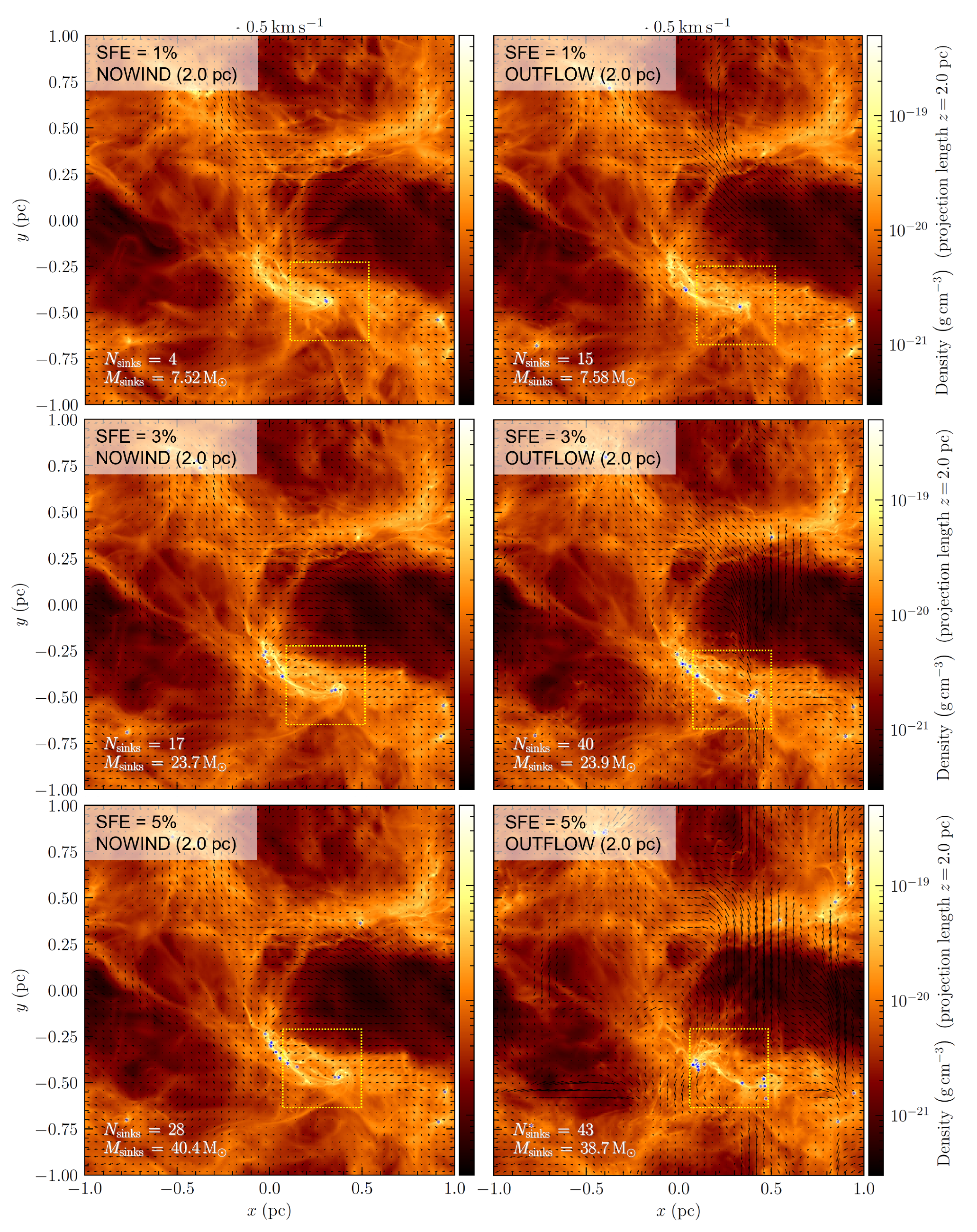}
	\caption{\label{fig:den20} Same as Fig.~\ref{fig:illu}, but here the gas column density is shown at SFE=1\% (top), 3\% (middle), and 5\% (bottom), in the NOWIND (left) and OUTFLOW simulation (right), respectively. The circular markers in each panel represent the position of the sink (star) particles formed in the simulations.}
\end{figure*}

Moreover, the fraction of compressive turbulence is an important aspect in the study of star formation. Compressive modes of the turbulence create density enhancements, which may serve as the birthplaces of stars \citep{2012ApJ...761..156F}. Therefore we decompose the structure function into longitudinal ${\rm SF_2^\parallel}(r)$ and transverse ${\rm SF_2^\perp}(r)$ components. The decomposition is performed in real space as
\begin{equation}
\begin{aligned}
        \delta \pmb{v} &= \pmb{v}(\pmb{x}+\pmb{r})-\pmb{v}(\pmb{x})\\
        {\rm SF_2^\parallel}(r)&=\langle\delta \pmb{v}^2(\hat{\delta \pmb{v}}\cdot\hat{\pmb{r}})\rangle_r\\
        {\rm SF_2^\perp}(r)&={\rm SF_2}(r)-{\rm SF_2^\parallel}(r).
\end{aligned}
\end{equation}
Accordingly, the fraction $\chi$ of the longitudinal velocity field mode $\chi(r)={\rm SF_2^\parallel}(r)/{\rm SF}_2(r)$.

\section{Results}
\label{sec:results}
\subsection{Basic evolution and structure of the clouds}
Fig.~\ref{fig:illu} shows maps of the projected gas density along the $z$-axis at two different evolutionary stages of the simulated cloud, i.e, at the stage of fully developed turbulence (left) and just before the $1^{\rm st}$ sink particle has formed (right). We can see a dense region developing towards the bottom right of the centre of the box, as indicated by the dotted square. These two stages are identical in the "NOWIND" and "OUTFLOW" simulations, because stars have not formed at this stage, and therefore, no outflows have been generated yet.

Starting from Fig.~\ref{fig:den20}, we examine how the evolution of the star formation is influenced by the outflow feedback. We show the projected gas density at three evolutionary stages, i.e., at star formation efficiencies SFE = 1\%, 3\%, and 5\%, which means 1\%, 3\%, and 5\% of the total mass of the cloud has formed stars, respectively. We observe that the outflows slow down the star-forming process, taking more time to reach the same SFE \citep[the reduction in the star formation rate due to outflow feedback is discussed in detail in][]{2021MNRAS.507.2448M}. For instance, the OUTFLOW model takes 0.07~Myr, 0.06~Myr, and 0.14~Myr longer to reach SFE = 1\%, 3\%, and 5\%, respectively, than the NOWIND model. These values vary for different turbulent seeds. On average, however, the OUTFLOW model generates more stars at the same SFE, with dense filaments breaking into several sub-fragments due to the action of the outflows. This phenomenon has been investigated in detail in \citet{2021MNRAS.507.2448M}. Possible explanations include that the outflow drives small-scale turbulence and enhances the fragmentation allowing the formation of more stars.


\begin{figure}
	\centering
	\includegraphics[width=1.0\linewidth]{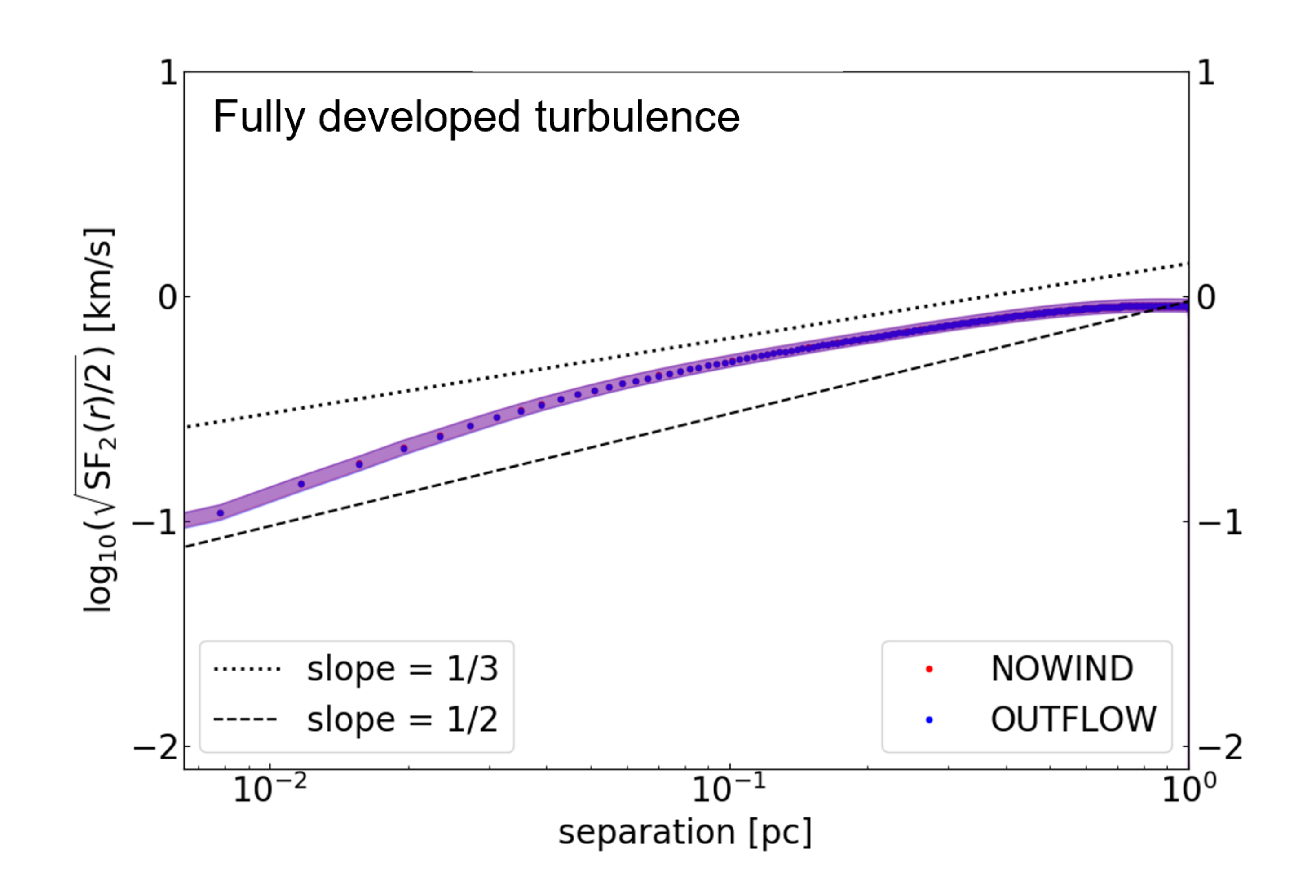}
	\caption{\label{fig:SF_full} Velocity dispersion as quantified by the square root of the normalized second order structure function (Eq.~\ref{eq:sfnorm}) as a function of separation (scale) covering the full simulation domain (2~pc) at the stage of fully developed turbulence. Since gravity and star formation have not started yet, the NOWIND (red) and OUTFLOW (blue) model are identical. The shaded areas represent the standard deviation over the set of 10~turbulent realisations per simulation model. To guide the eye, the dotted and dashed lines represent power-law slopes of $1/3$ and $1/2$, for comparison with a Kolmogorov and Burgers scaling of turbulence, respectively. On the cloud scale (1~pc), we see that the 2nd-order SF reflects the input velocity dispersion of 1~km/s, as set by the turbulence driving.}
\end{figure}

\subsection{The structure function on cloud scales}

\begin{figure*}
	\centering
	\includegraphics[width=1.0\linewidth]{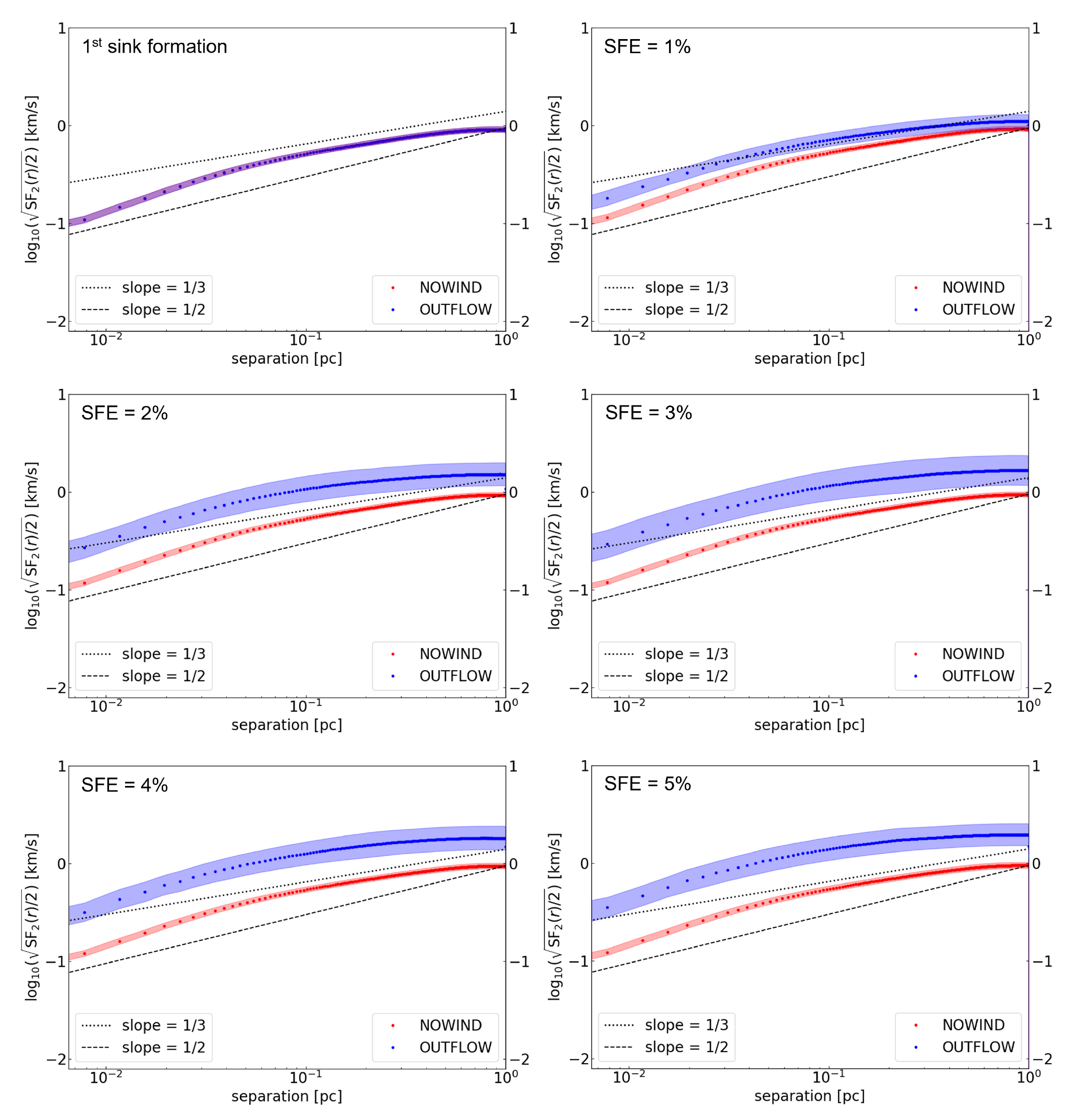}
	\caption{\label{fig:SF20} Velocity dispersion as quantified by the square root of the normalized second order structure function (Eq.~\ref{eq:sfnorm}) as a function of separation (scale) covering the full simulation domain (2~pc). We compare the NOWIND (red) and OUTFLOW models (blue) at six snapshots, as indicated in the top left of each panel: after 1st sink particle formation, and at $\mathrm{SFE}=1$, 2, 3, 4, 5\%, respectively. The shaded areas represent the standard deviation over the set of 10~turbulent realisations per simulation model. To guide the eye, the dotted lines represent power-law slopes of $1/3$ and $1/2$, for comparison with a Kolmogorov and Burgers scaling of turbulence, respectively.}
\end{figure*}

In order to quantify the distribution of turbulent motions on different scales of the cloud, we show in Fig.~\ref{fig:SF20} the square root of the normalized 2$^{\rm nd}$ order structure function, which is equivalent to the 3D velocity dispersion $\delta v(r)$, as defined by Eq.~(\ref{eq:sfnorm}). All structure functions have been averaged over ten simulation realizations of the turbulence, for each model, NOWIND and OUTFLOW. Before star formation begins and outflows are launched (snapshot just before $1^{st}$ sink formation), the structure functions are identical for the NOWIND and OUTFLOW models, representing a state of fully developed turbulence with gravity having modified the system such that local star formation is imminent. The SF at the state of of fully developed turbulence in presented in Fig.~\ref{fig:SF_full}. The dotted and dashed lines indicate a power-law scaling with a slope of $1/3$ \citep{Kolmogorov1941c} and $1/2$ \citep{Burgers1948}, respectively. On scales of $0.05\lesssim r/\mathrm{pc}\lesssim0.5$, the simulations roughly follow a scaling in between Kolmogorov and Burgers turbulence, as expected for a mildly supersonic turbulent medium \citep{2010A&A...512A..81F,Federrath2013,2021NatAs...5..365F}. Such scaling of velocity fluctuation is widely observed in diffuse neutral hydrogen clouds and dense molecular clouds \citep{2008ApJ...680..420H,2009SSRv..143..357L}. On scales $r\lesssim0.05$~pc, the velocity dispersion drops due to numerical dissipation starting to act on scales below $\sim30$~grid cells \citep{KitsionasEtAl2009,FederrathEtAl2011}. Although the base grid (lowest level grid, covering the entire domain) has a resolution of only $256$ cells in each cartesian direction, the simulations use AMR with Jeans refinement, up to a maximum effective resolution of $4096$ cells in each spatial direction.

Differences in the structure functions appear when stars form. While the velocity dispersion in the NOWIND model does not change significantly in time, the OUTFLOW model clearly evolves over time, such that the velocity dispersion increases until $\mathrm{SFE}\sim2$--$3\%$, i.e., the amplitude of the structure function grows. After $\mathrm{SFE}\sim2$--$3\%$, the velocity dispersion appears to reach a steady state, in which the dissipation of turbulent flows is balanced by the injection of energy from the outflows, at a roughly constant rate. This is consistent with the initial rise in the SFR observed in these models, followed by a steady-state (nearly constant) SFR \citep{2021MNRAS.507.2448M}. The rise in velocity dispersion in the OUTFLOW model is caused by the injected energy of the jets/outflows, driving turbulence from small scales, i.e., the jets originate locally around each protostar, on scales as small as sub-AU \citep{FederrathEtAl2014}. However, as the jets propagate through the ambient parental cloud from which they formed, their influence and energy spreads to a scale $\gtrsim1$~pc. Consequently, the velocity dispersion in the presence of outflow feedback increases over all resolvable scales up to $\sim1$~pc.

We also see that the OUTFLOW model exhibits a somewhat more curved structure function, i.e., the power-law scaling range appears to be suppressed; or at there is no clear power-law scaling range seen anymore, especially at later times when the influence of the outflows grows. The absence of the scaling range may be due to limited resolution or due to the driving effect of the outflows, mixing into the large-scale turbulent 'cascade'. The NOWIND model, however, exhibits insignificant changes in the structure function, in terms of either amplitude or scaling. This can be understood based on the fact that the gravitational collapse happens on relatively small scales, and averaging the structure function over the full domain, including diffuse regions, means that the effect of gravity is hardly noticeable in the cloud-scale structure functions. Therefore, we zoom in onto a dense star-forming clump in the following section, in order to determine the effects of gravity and outflows on the dense, star-forming regions of the cloud.

\subsection{The structure function of dense, star-forming regions}
\label{subsec: zoomSF}
\begin{figure*}
	\centering
	\includegraphics[width=0.99\linewidth]{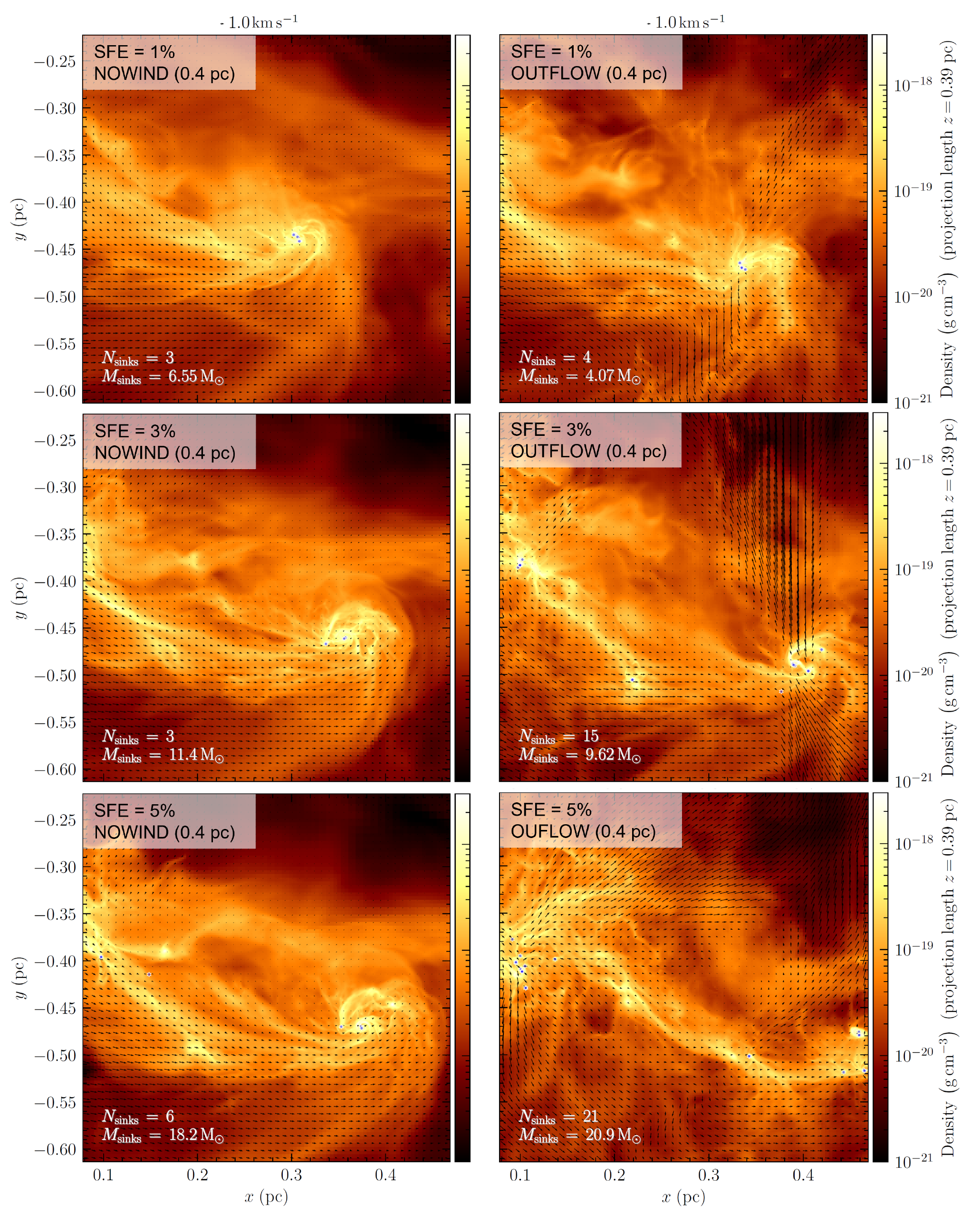}
	\caption{\label{fig:den04} Same as Fig.~\ref{fig:den20}, but for a high-density, star-forming region of size 0.4~pc.}
\end{figure*}

\begin{figure*}
	\centering
	\includegraphics[width=1.0\linewidth]{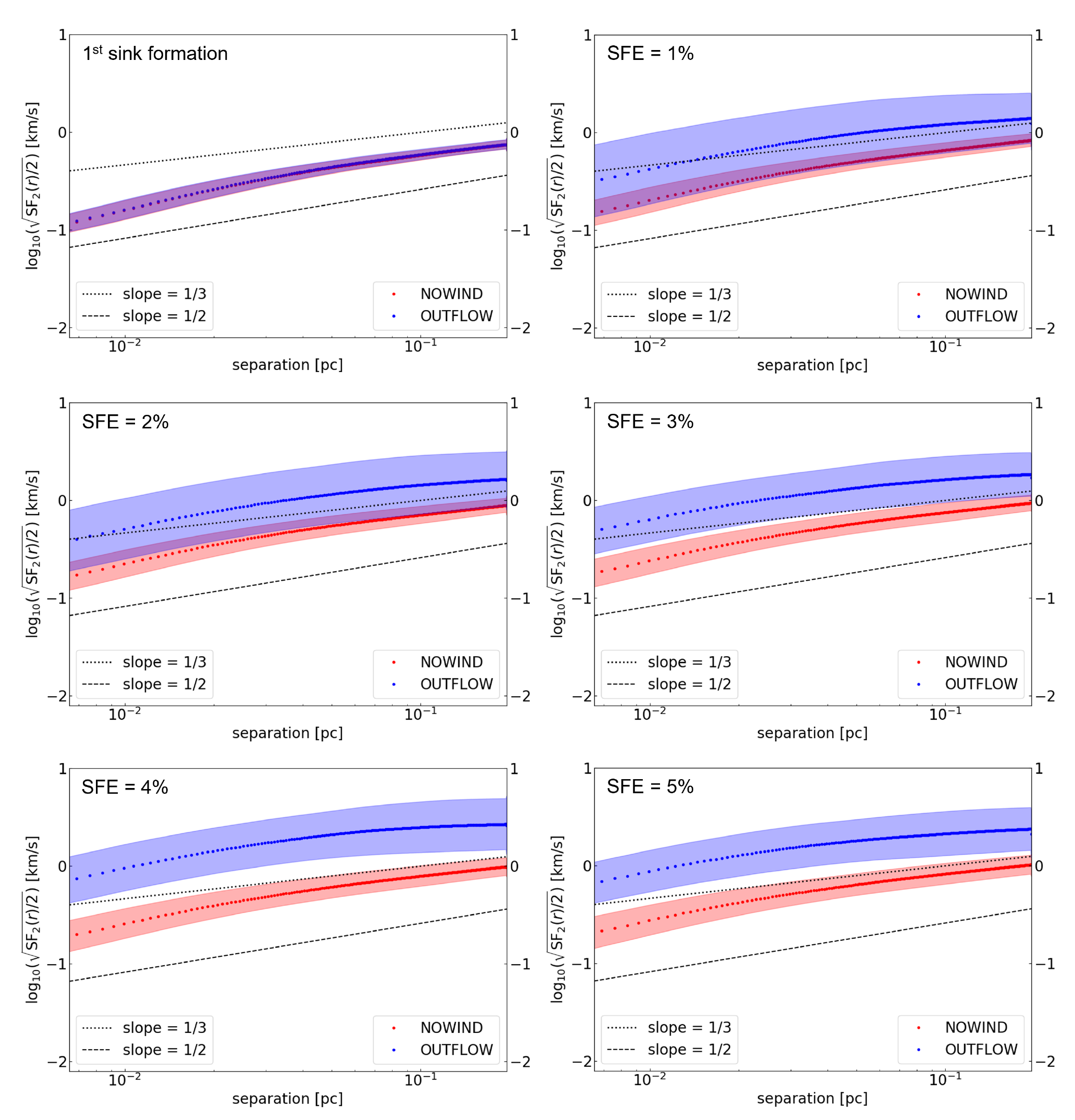}
	\caption{\label{fig:SF04} Same as Fig.~\ref{fig:SF20}, but for the zoom-in region outlined in Fig.~\ref{fig:den20}.}
\end{figure*}
\begin{figure*}
	\centering
	\includegraphics[width=0.92\linewidth]{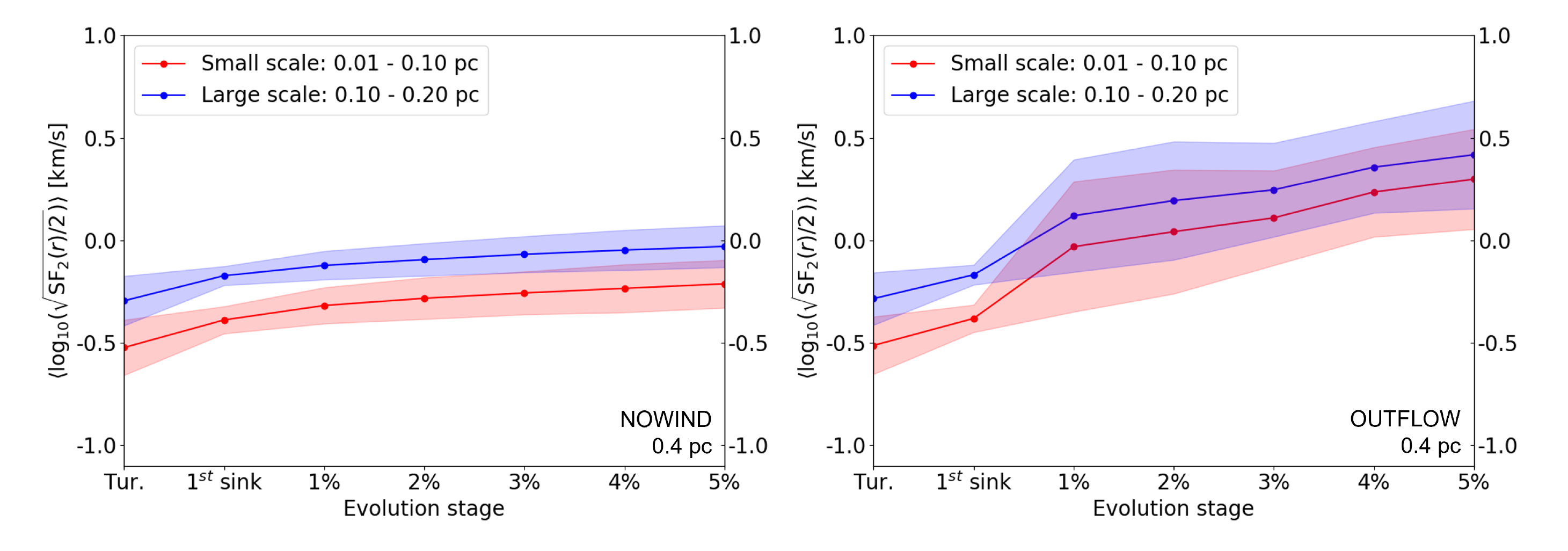}
	\caption{\label{fig:SF_time04} The plots show the square root of the normalized second order structure function as a function of time. The structure function is averaged over the small scale range (0.01 - 0.10 pc; red) and large scale range (0.10 - 0.20 pc; blue) in the zoom-in 0.4 pc region. Two physical models: "NOWIND" (left) and "OUTFLOW" (right) are presented here. The coloured shallows represent the standard deviation over the set of 10 simulations.}
\end{figure*}

\begin{figure*}
	\centering
	\includegraphics[width=0.92\linewidth]{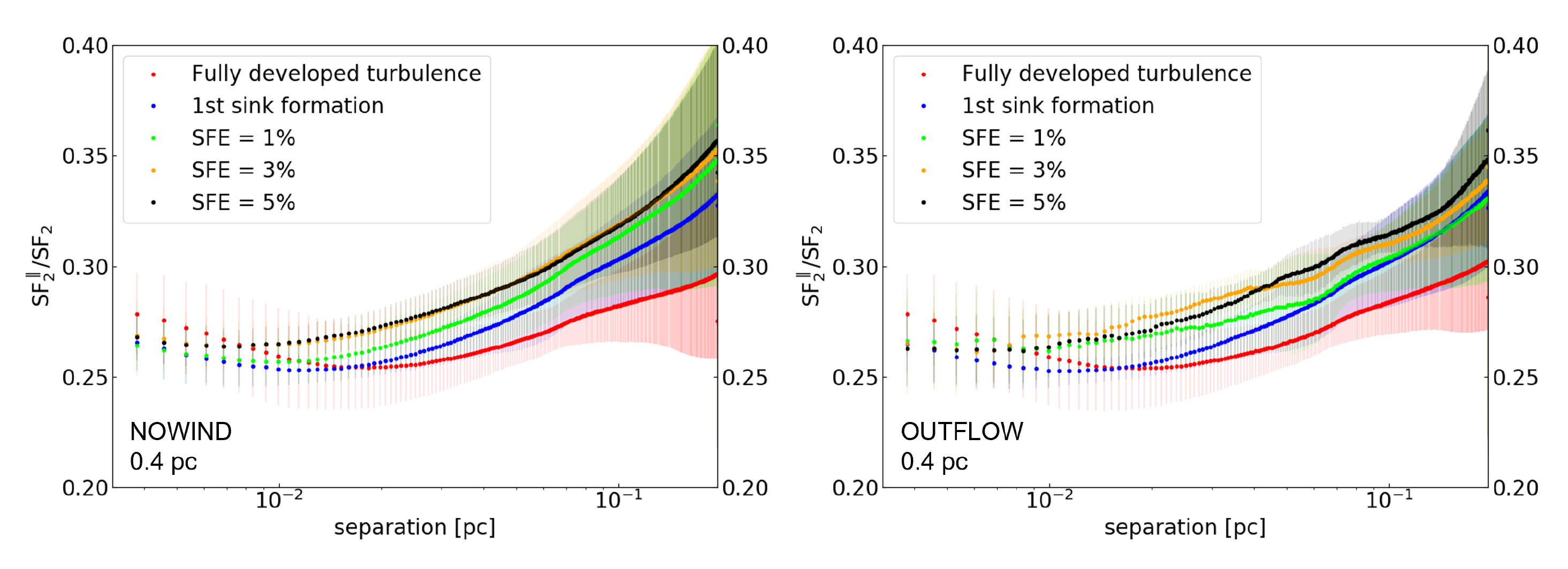}
	\caption{\label{fig:frac_long04} The plots present the energy fraction of longitudinal component of the structure function as a function of separation. Two physical models: "NOWIND" (left) and "OUTFLOW" (right) are presented here. The error bar is by the standard deviation over the set of 10 simulations.}
\end{figure*}

\begin{figure*}
	\centering
	\includegraphics[width=0.92\linewidth]{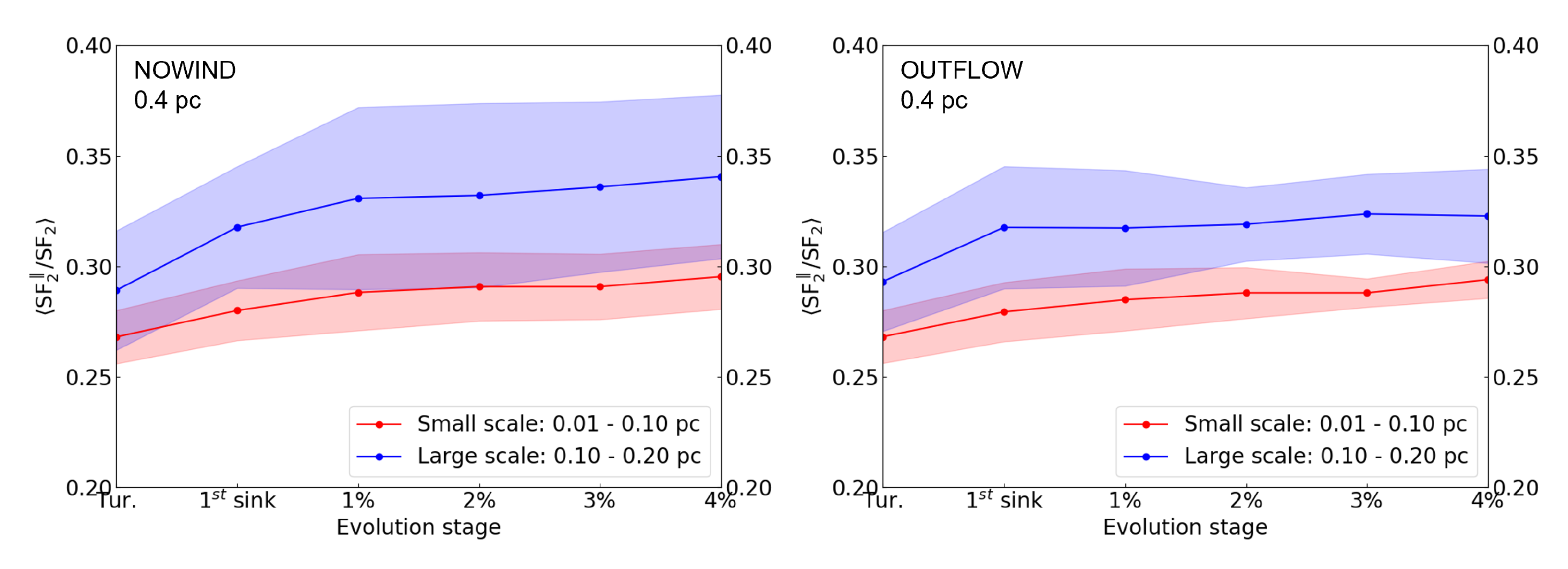}
	\caption{\label{fig:Efrac04}Same as Fig.~\ref{fig:SF_time04}, but for the energy fraction of the longitudinal structure function.}
\end{figure*}

Fig.~\ref{fig:den04} presents the time evolution of a 0.4~pc zoom-in region in the cloud of NOWIND and OUTFLOW models with turbulence realization T2. The zoom-in is selected by searching for the maximum density point in 3D space at the evolutionary stage SFE = 5\%. Consequently, its spatial position is fixed among all stages. Similar to the case of the full simulation plotted in Fig.~\ref{fig:den20}, the OUTFLOW model produces a higher number of stars and the difference between NOWIND and OUTFLOW is more significant here. We can see the fragmentation in the OUTFLOW model is stronger with several sub-filamentary structures. In particular, two extra collapsing centers appear in the stages SFE = 3\% and SFE = 5\%. This suggests that with identical initial conditions protostellar outflows can trigger local star formation in nearby filaments. Note that outflows create new density perturbations at small scales, which increases gravo-turbulent fragmentation \citep{2021MNRAS.507.2448M}.

Moreover, in Fig.~\ref{fig:SF04}, we show the velocity dispersion (i.e., the square root of the normalized 2$^{\rm nd}$ order structure-function) as a function of separation. Compared with the full box cases (see Fig.~\ref{fig:SF20}), the velocity dispersion increases in both NOWIND and OUTFLOW models. The increment seen in the NOWIND model is contributed by the effect of gravity, including the infall velocity and probably gravity-driven turbulence. This effect focuses on small scales and therefore is more apparent in the zoom-in dense region, especially at the SFE = 5\% stage. The velocity dispersion is further signified in the OUTFLOW models. Here the dispersion includes the contribution from the original turbulence, and turbulence produced by gravitational infall and by the outflows. Compared with gravity, the influence of the outflows seems to extend to larger scales (see Fig.~\ref{fig:SF20}). Despite the OUTFLOW models' velocity dispersion getting slightly curved, the overall scaling is still approximately in between Kolmogorov (i.e., $\sigma_v(r)\propto r^{1/3}$) and Burgers (i.e., $\sigma_v(r) \propto r^{1/2}$) turbulence.

In Fig.~\ref{fig:SF_time04}, we investigate how the velocity dispersion evolves as a function of time. Here we define the separation in the range of 0.01 - 0.10~pc as small scale and 0.10 - 0.20~pc as large scale. The velocity dispersion is averaged over these two scales, respectively. For both NOWIND and OUTFLOW models, the dispersion increases with time as gravity and outflows get stronger. Gravity amplifies the velocity dispersion by a factor of $\sim2$ approximately from the beginning (fully developed turbulence) to SFE = 5\%. However, with outflow feedback, the velocity dispersion increases by up to a factor of $\sim7$ from the the time of fully-developed turbulence to SFE = 5\%, for both small and large scales. In particular, we see that the OUTFLOW model exhibits a larger standard deviation, which is quantified by the differences over the ten realizations of the simulations. 
This suggests that compared with gravity, outflows could perturb the velocity field further. The effect of the outflows is more sensitive to the initial turbulent seed.
 
Finally, we conclude that both gravity and outflow feedback affect the velocity field, driving larger velocity dispersion in star-forming regions. In particular, outflows enhance fragmentation \citep{FederrathEtAl2014,2021MNRAS.507.2448M,2022arXiv220100882G}. Compared with gravity, the outflow has its effect on larger scales up to $\sim1$~pc at least (see Fig.~\ref{fig:SF20}) and is more sensitive to the initial turbulent seeds.

To observationally test our results, it is important to obtain information of velocity fluctuations. Using Doppler-shifted emission lines, this is achievable via the Velocity Channel Analysis (VCA; \citealt{2000ApJ...537..720L}), the Velocity Coordinate Spectrum (VCS; \citealt{2004JKAS...37..563L}), and the Principal Component Analysis (PCA; \citealt{2008ApJ...680..420H}). It is also possible to infer some of the gas turbulence properties on cloud scales using the velocity of young stellar objects \citep{2021ApJ...907L..40H}.
\subsection{The energy fraction of the longitudinal velocity component}
Fig.~\ref{fig:frac_long04} presents the energy fraction of the longitudinal velocity component as a function of separation. Initially, the turbulence is driven by the mix of compressive (i.e., longitudinal) and solenoidal (i.e., transverse) modes. The fraction of the longitudinal component in the turbulence driving is constructed to be 1/3 at driving scale (referred to as the 'mixed' driving mode). Before gravity starts acting, i.e., at the stage of fully developed turbulence, the longitudinal fraction decreases from the scale of 0.2~pc to 0.02~pc. It suggests that when turbulence cascades to small scales, the compressive component dissipates somewhat faster than the solenoidal component \citep{2016ApJ...822...11P}. Gravity, in addition, increases the longitudinal fraction by generating longitudinal velocity fields due to infall and gravity-driven turbulence. A portion of the gravitational potential is converted to kinetic energy in this process. This increment is the most significant from the stage of fully developed turbulence to the 1st sink formation stage. However, after SFE = 1\%, the fraction's increment rate gets slower. One possibility is that the zoom-in region may evolve slower than the global simulation so that the local SFE in the subregion varies significantly between different realizations. It is also likely that the amplification of longitudinal modes is mainly contributed by the cloud-scale ($\sim1$~pc) contraction. At later stages, stars start forming locally so that the longitudinal field exhibits only relatively small changes when looked at from a global perspective, i.e., when averaged over the total volume; however, close to the stars, the longitudinal component increases. The global fraction of longitudinal modes finally reaches $\sim0.35$ at 0.2~pc scales on average, with a maximum of $\sim 0.40$. Compared with the NOWIND model, the OUTFLOW model shows a similar trend, but the increment after the 1st sink formation stage is suppressed so that its maximum value is smaller than 0.40. This is because outflow feedback slows down star formation and therefore slows down the development of compressible modes caused by collapse. Another contributor to reducing the relative amount of compressible modes in the OUTFLOW case is that outflows may drive solenoidal modes, therefore enhancing their relative contribution.

To study the time evolution of the longitudinal energy fraction, as before (cf.~Fig.~\ref{fig:frac_long04}, we distinguish two scale ranges: (i) small scale from 0.01 to 0.10~pc and (ii) large scale from 0.10 to 0.20~pc. As shown in Fig.~\ref{fig:Efrac04}, the longitudinal fraction keeps increasing with time. Large scales host more longitudinal modes with an average longitudinal fraction of $\sim0.32$, while small scales exhibit a fraction of $\sim0.28$. Note that the fraction is averaged over the defined scale ranges. In this case, the OUTFLOW model is similar to the NOWIND case, but exhibits smaller values of the fraction at large scales. This suggests that the suppression of the longitudinal velocity field's fraction by outflows is most significant on large scales. It leads to a reduction in the star formation rate compared to when outflow feedback is not included.

\section{Summary} 
\label{sec:con}
In this work, we use three-dimensional MHD simulations of star cluster formation to study the interplay of turbulence, magnetic fields, self-gravity, radiation, and stellar feedback. We focus on analyzing the effects of self-gravity and stellar feedback on the second-order structure-function of turbulent velocities. The stellar feedback slightly changes the scaling of velocity fluctuations but the overall scaling is still in between Kolmogorov $\sigma_v(r) \propto r^{1/3}$ and Burgers $\sigma_v(r) \propto r^{1/2}$ type. We observe that both self-gravity and protostellar outflows increase the turbulent velocity dispersion. In particular, the outflows increase fragmentation and amplify the velocity fluctuations by a factor of $\sim7$. The effect of the outflows spans from stellar/disk scales to cloud-scales $\gtrsim1$~pc. The amplified velocity fluctuations may provide more support against gravity and enhance fragmentation on small scale. The role of self-gravity is more significant on the scales of dense clumps and it increases the fraction of the compressive velocity component. On the other hand, the effect of the outflows is to slightly reduce the relative fraction of compressible velocity modes on scales $\gtrsim0.1$~pc. Thus, outflows primarily drive solenoidal modes. Overall, the compressive velocity fraction stays close to the natural mixture of $\sim1/3$, but is systematically smaller on small scales, compared to large scales, which suggests that the compressive velocity component dissipates energy faster on small scales.

\section*{Acknowledgements}
C.F.~acknowledges funding provided by the Australian Research Council (Future Fellowship FT180100495), and the Australia-Germany Joint Research Cooperation Scheme (UA-DAAD). S.X.~acknowledges the support for this work provided by NASA through the NASA Hubble Fellowship grant \# HST-HF2-51473.001-A awarded by the Space Telescope Science Institute, which is operated by the Association of Universities for Research in Astronomy, Incorporated, under NASA contract NAS5-26555. We further acknowledge high-performance computing resources provided by the Australian National Computational Infrastructure (grant~ek9) in the framework of the National Computational Merit Allocation Scheme and the ANU Merit Allocation Scheme, and by the Leibniz Rechenzentrum and the Gauss Centre for Supercomputing (grant~pr32lo and GCS Large-scale project~10391). The simulation software FLASH was in part developed by the DOE-supported Flash Center for Computational Science.

\section*{Data Availability}
The data underlying this article will be shared on reasonable request to the corresponding author. 


\bibliographystyle{mnras}
\bibliography{example} 

\begin{thebibliography}{}
\makeatletter
\relax
\def\mn@urlcharsother{\let\do\@makeother \do\$\do\&\do\#\do\^\do\_\do\%\do\~}
\def\mn@doi{\begingroup\mn@urlcharsother \@ifnextchar [ {\mn@doi@}
  {\mn@doi@[]}}
\def\mn@doi@[#1]#2{\def\@tempa{#1}\ifx\@tempa\@empty \href
  {http://dx.doi.org/#2} {doi:#2}\else \href {http://dx.doi.org/#2} {#1}\fi
  \endgroup}
\def\mn@eprint#1#2{\mn@eprint@#1:#2::\@nil}
\def\mn@eprint@arXiv#1{\href {http://arxiv.org/abs/#1} {{\tt arXiv:#1}}}
\def\mn@eprint@dblp#1{\href {http://dblp.uni-trier.de/rec/bibtex/#1.xml}
  {dblp:#1}}
\def\mn@eprint@#1:#2:#3:#4\@nil{\def\@tempa {#1}\def\@tempb {#2}\def\@tempc
  {#3}\ifx \@tempc \@empty \let \@tempc \@tempb \let \@tempb \@tempa \fi \ifx
  \@tempb \@empty \def\@tempb {arXiv}\fi \@ifundefined
  {mn@eprint@\@tempb}{\@tempb:\@tempc}{\expandafter \expandafter \csname
  mn@eprint@\@tempb\endcsname \expandafter{\@tempc}}}

\bibitem[\protect\citeauthoryear{{Armstrong}, {Rickett}  \&
  {Spangler}}{{Armstrong} et~al.}{1995}]{1995ApJ...443..209A}
{Armstrong} J.~W.,  {Rickett} B.~J.,   {Spangler} S.~R.,  1995, \mn@doi [\apj]
  {10.1086/175515}, \href
  {https://ui.adsabs.harvard.edu/abs/1995ApJ...443..209A} {443, 209}

\bibitem[\protect\citeauthoryear{{Berger} \& {Colella}}{{Berger} \&
  {Colella}}{1989}]{1989JCoPh..82...64B}
{Berger} M.~J.,  {Colella} P.,  1989, \mn@doi [Journal of Computational
  Physics] {10.1016/0021-9991(89)90035-1}, \href
  {https://ui.adsabs.harvard.edu/abs/1989JCoPh..82...64B} {82, 64}

\bibitem[\protect\citeauthoryear{Burgers}{Burgers}{1948}]{Burgers1948}
Burgers J.~M.,  1948, in , Vol.~1, Advances in applied mechanics.
Elsevier, pp 171--199

\bibitem[\protect\citeauthoryear{{Burkhart}, {Collins}  \&
  {Lazarian}}{{Burkhart} et~al.}{2015}]{2015ApJ...808...48B}
{Burkhart} B.,  {Collins} D.~C.,   {Lazarian} A.,  2015, \mn@doi [\apj]
  {10.1088/0004-637X/808/1/48}, \href
  {https://ui.adsabs.harvard.edu/abs/2015ApJ...808...48B} {808, 48}

\bibitem[\protect\citeauthoryear{{Carroll}, {Frank}, {Blackman}, {Cunningham}
  \& {Quillen}}{{Carroll} et~al.}{2009}]{2009ApJ...695.1376C}
{Carroll} J.~J.,  {Frank} A.,  {Blackman} E.~G.,  {Cunningham} A.~J.,
  {Quillen} A.~C.,  2009, \mn@doi [\apj] {10.1088/0004-637X/695/2/1376}, \href
  {https://ui.adsabs.harvard.edu/abs/2009ApJ...695.1376C} {695, 1376}

\bibitem[\protect\citeauthoryear{{Chepurnov} \& {Lazarian}}{{Chepurnov} \&
  {Lazarian}}{2010}]{2010ApJ...710..853C}
{Chepurnov} A.,  {Lazarian} A.,  2010, \mn@doi [\apj]
  {10.1088/0004-637X/710/1/853}, \href
  {https://ui.adsabs.harvard.edu/abs/2010ApJ...710..853C} {710, 853}

\bibitem[\protect\citeauthoryear{{Chira}, {Ib{\'a}{\~n}ez-Mej{\'\i}a}, {Mac
  Low}  \& {Henning}}{{Chira} et~al.}{2019}]{2019A&A...630A..97C}
{Chira} R.~A.,  {Ib{\'a}{\~n}ez-Mej{\'\i}a} J.~C.,  {Mac Low} M.~M.,
  {Henning} T.,  2019, \mn@doi [\aap] {10.1051/0004-6361/201833970}, \href
  {https://ui.adsabs.harvard.edu/abs/2019A&A...630A..97C} {630, A97}

\bibitem[\protect\citeauthoryear{{Cho} \& {Lazarian}}{{Cho} \&
  {Lazarian}}{2003}]{2003MNRAS.345..325C}
{Cho} J.,  {Lazarian} A.,  2003, \mn@doi [\mnras]
  {10.1046/j.1365-8711.2003.06941.x}, \href
  {https://ui.adsabs.harvard.edu/abs/2003MNRAS.345..325C} {345, 325}

\bibitem[\protect\citeauthoryear{{Dubey} et~al.,}{{Dubey}
  et~al.}{2008}]{2008ASPC..385..145D}
{Dubey} A.,  et~al., 2008, in {Pogorelov} N.~V.,  {Audit} E.,   {Zank} G.~P.,
  eds,  Astronomical Society of the Pacific Conference Series Vol. 385,
  Numerical Modeling of Space Plasma Flows. p.~145

\bibitem[\protect\citeauthoryear{{Elmegreen}}{{Elmegreen}}{1993}]{1993ApJ...419L..29E}
{Elmegreen} B.~G.,  1993, \mn@doi [\apjl] {10.1086/187129}, \href
  {https://ui.adsabs.harvard.edu/abs/1993ApJ...419L..29E} {419, L29}

\bibitem[\protect\citeauthoryear{{Esquivel}, {Lazarian}, {Pogosyan}  \&
  {Cho}}{{Esquivel} et~al.}{2003}]{2003MNRAS.342..325E}
{Esquivel} A.,  {Lazarian} A.,  {Pogosyan} D.,   {Cho} J.,  2003, \mn@doi
  [\mnras] {10.1046/j.1365-8711.2003.06551.x}, \href
  {https://ui.adsabs.harvard.edu/abs/2003MNRAS.342..325E} {342, 325}

\bibitem[\protect\citeauthoryear{{Eswaran} \& {Pope}}{{Eswaran} \&
  {Pope}}{1988}]{1988CF.....16..257E}
{Eswaran} V.,  {Pope} S.~B.,  1988, Computers and Fluids, \href
  {https://ui.adsabs.harvard.edu/abs/1988CF.....16..257E} {16, 257}

\bibitem[\protect\citeauthoryear{{Ewertowski} \& {Basu}}{{Ewertowski} \&
  {Basu}}{2013}]{2013ApJ...767...33E}
{Ewertowski} B.,  {Basu} S.,  2013, \mn@doi [\apj]
  {10.1088/0004-637X/767/1/33}, \href
  {https://ui.adsabs.harvard.edu/abs/2013ApJ...767...33E} {767, 33}

\bibitem[\protect\citeauthoryear{{Federrath}}{{Federrath}}{2013}]{Federrath2013}
{Federrath} C.,  2013, \mn@doi [\mnras] {10.1093/mnras/stt1644}, \href
  {https://ui.adsabs.harvard.edu/abs/2013MNRAS.436.1245F} {436, 1245}

\bibitem[\protect\citeauthoryear{{Federrath}}{{Federrath}}{2015}]{2015MNRAS.450.4035F}
{Federrath} C.,  2015, \mn@doi [\mnras] {10.1093/mnras/stv941}, \href
  {https://ui.adsabs.harvard.edu/abs/2015MNRAS.450.4035F} {450, 4035}

\bibitem[\protect\citeauthoryear{{Federrath} \& {Klessen}}{{Federrath} \&
  {Klessen}}{2012}]{2012ApJ...761..156F}
{Federrath} C.,  {Klessen} R.~S.,  2012, \mn@doi [\apj]
  {10.1088/0004-637X/761/2/156}, \href
  {https://ui.adsabs.harvard.edu/abs/2012ApJ...761..156F} {761, 156}

\bibitem[\protect\citeauthoryear{{Federrath} \& {Klessen}}{{Federrath} \&
  {Klessen}}{2013}]{FederrathKlessen2013}
{Federrath} C.,  {Klessen} R.~S.,  2013, \mn@doi [\apj]
  {10.1088/0004-637X/763/1/51}, \href
  {https://ui.adsabs.harvard.edu/abs/2013ApJ...763...51F} {763, 51}

\bibitem[\protect\citeauthoryear{{Federrath}, {Klessen}  \&
  {Schmidt}}{{Federrath} et~al.}{2008}]{FederrathKlessenSchmidt2008}
{Federrath} C.,  {Klessen} R.~S.,   {Schmidt} W.,  2008, \mn@doi [\apjl]
  {10.1086/595280}, \href
  {https://ui.adsabs.harvard.edu/abs/2008ApJ...688L..79F} {688, L79}

\bibitem[\protect\citeauthoryear{{Federrath}, {Klessen}  \&
  {Schmidt}}{{Federrath} et~al.}{2009}]{FederrathEtAl2009}
{Federrath} C.,  {Klessen} R.~S.,   {Schmidt} W.,  2009, \mn@doi [\apj]
  {10.1088/0004-637X/692/1/364}, \href
  {https://ui.adsabs.harvard.edu/abs/2009ApJ...692..364F} {692, 364}

\bibitem[\protect\citeauthoryear{{Federrath}, {Roman-Duval}, {Klessen},
  {Schmidt}  \& {Mac Low}}{{Federrath} et~al.}{2010}]{2010A&A...512A..81F}
{Federrath} C.,  {Roman-Duval} J.,  {Klessen} R.~S.,  {Schmidt} W.,   {Mac Low}
  M.~M.,  2010, \mn@doi [\aap] {10.1051/0004-6361/200912437}, \href
  {https://ui.adsabs.harvard.edu/abs/2010A&A...512A..81F} {512, A81}

\bibitem[\protect\citeauthoryear{{Federrath}, {Chabrier}, {Schober},
  {Banerjee}, {Klessen}  \& {Schleicher}}{{Federrath}
  et~al.}{2011}]{FederrathEtAl2011}
{Federrath} C.,  {Chabrier} G.,  {Schober} J.,  {Banerjee} R.,  {Klessen}
  R.~S.,   {Schleicher} D.~R.~G.,  2011, \mn@doi [\prl]
  {10.1103/PhysRevLett.107.114504}, \href
  {https://ui.adsabs.harvard.edu/abs/2011PhRvL.107k4504F} {107, 114504}

\bibitem[\protect\citeauthoryear{{Federrath}, {Schr{\"o}n}, {Banerjee}  \&
  {Klessen}}{{Federrath} et~al.}{2014a}]{2014ApJ...790..128F}
{Federrath} C.,  {Schr{\"o}n} M.,  {Banerjee} R.,   {Klessen} R.~S.,  2014a,
  \mn@doi [\apj] {10.1088/0004-637X/790/2/128}, \href
  {https://ui.adsabs.harvard.edu/abs/2014ApJ...790..128F} {790, 128}

\bibitem[\protect\citeauthoryear{{Federrath}, {Schr{\"o}n}, {Banerjee}  \&
  {Klessen}}{{Federrath} et~al.}{2014b}]{FederrathEtAl2014}
{Federrath} C.,  {Schr{\"o}n} M.,  {Banerjee} R.,   {Klessen} R.~S.,  2014b,
  \mn@doi [\apj] {10.1088/0004-637X/790/2/128}, \href
  {https://ui.adsabs.harvard.edu/abs/2014ApJ...790..128F} {790, 128}

\bibitem[\protect\citeauthoryear{{Federrath}, {Klessen}, {Iapichino}  \&
  {Beattie}}{{Federrath} et~al.}{2021}]{2021NatAs...5..365F}
{Federrath} C.,  {Klessen} R.~S.,  {Iapichino} L.,   {Beattie} J.~R.,  2021,
  \mn@doi [Nature Astronomy] {10.1038/s41550-020-01282-z}, \href
  {https://ui.adsabs.harvard.edu/abs/2021NatAs...5..365F} {5, 365}

\bibitem[\protect\citeauthoryear{{Ferri{\`e}re}}{{Ferri{\`e}re}}{2001}]{2001RvMP...73.1031F}
{Ferri{\`e}re} K.~M.,  2001, \mn@doi [Reviews of Modern Physics]
  {10.1103/RevModPhys.73.1031}, \href
  {https://ui.adsabs.harvard.edu/abs/2001RvMP...73.1031F} {73, 1031}

\bibitem[\protect\citeauthoryear{{Ferri{\`e}re}}{{Ferri{\`e}re}}{2020}]{2020PPCF...62a4014F}
{Ferri{\`e}re} K.,  2020, \mn@doi [Plasma Physics and Controlled Fusion]
  {10.1088/1361-6587/ab49eb}, \href
  {https://ui.adsabs.harvard.edu/abs/2020PPCF...62a4014F} {62, 014014}

\bibitem[\protect\citeauthoryear{{Fryxell} et~al.,}{{Fryxell}
  et~al.}{2000}]{2000ApJS..131..273F}
{Fryxell} B.,  et~al., 2000, \mn@doi [\apjs] {10.1086/317361}, \href
  {https://ui.adsabs.harvard.edu/abs/2000ApJS..131..273F} {131, 273}

\bibitem[\protect\citeauthoryear{{Grudi{\'c}}, {Guszejnov}, {Hopkins}, {Offner}
   \& {Faucher-Gigu{\`e}re}}{{Grudi{\'c}} et~al.}{2021}]{GrudicEtAl2021}
{Grudi{\'c}} M.~Y.,  {Guszejnov} D.,  {Hopkins} P.~F.,  {Offner} S. S.~R.,
  {Faucher-Gigu{\`e}re} C.-A.,  2021, \mn@doi [\mnras]
  {10.1093/mnras/stab1347}, \href
  {https://ui.adsabs.harvard.edu/abs/2021MNRAS.506.2199G} {506, 2199}

\bibitem[\protect\citeauthoryear{{Grudi{\'c}}, {Guszejnov}, {Offner}, {Rosen},
  {Raju}, {Faucher-Gigu{\`e}re}  \& {Hopkins}}{{Grudi{\'c}}
  et~al.}{2022}]{2022arXiv220100882G}
{Grudi{\'c}} M.~Y.,  {Guszejnov} D.,  {Offner} S. S.~R.,  {Rosen} A.~L.,
  {Raju} A.~N.,  {Faucher-Gigu{\`e}re} C.-A.,   {Hopkins} P.~F.,  2022, arXiv
  e-prints, \href {https://ui.adsabs.harvard.edu/abs/2022arXiv220100882G} {p.
  arXiv:2201.00882}

\bibitem[\protect\citeauthoryear{{Ha}, {Li}, {Xu}, {Kounkel}  \& {Li}}{{Ha}
  et~al.}{2021}]{2021ApJ...907L..40H}
{Ha} T.,  {Li} Y.,  {Xu} S.,  {Kounkel} M.,   {Li} H.,  2021, \mn@doi [\apjl]
  {10.3847/2041-8213/abd8c9}, \href
  {https://ui.adsabs.harvard.edu/abs/2021ApJ...907L..40H} {907, L40}

\bibitem[\protect\citeauthoryear{{Hansen}, {McKee}  \& {Klein}}{{Hansen}
  et~al.}{2011}]{2011ApJ...738...88H}
{Hansen} C.~E.,  {McKee} C.~F.,   {Klein} R.~I.,  2011, \mn@doi [\apj]
  {10.1088/0004-637X/738/1/88}, \href
  {https://ui.adsabs.harvard.edu/abs/2011ApJ...738...88H} {738, 88}

\bibitem[\protect\citeauthoryear{{Heyer} \& {Brunt}}{{Heyer} \&
  {Brunt}}{2004}]{2004ApJ...615L..45H}
{Heyer} M.~H.,  {Brunt} C.~M.,  2004, \mn@doi [\apjl] {10.1086/425978}, \href
  {https://ui.adsabs.harvard.edu/abs/2004ApJ...615L..45H} {615, L45}

\bibitem[\protect\citeauthoryear{{Heyer}, {Gong}, {Ostriker}  \&
  {Brunt}}{{Heyer} et~al.}{2008}]{2008ApJ...680..420H}
{Heyer} M.,  {Gong} H.,  {Ostriker} E.,   {Brunt} C.,  2008, \mn@doi [\apj]
  {10.1086/587510}, \href
  {https://ui.adsabs.harvard.edu/abs/2008ApJ...680..420H} {680, 420}

\bibitem[\protect\citeauthoryear{{Hu}, {Lazarian}  \& {Yuen}}{{Hu}
  et~al.}{2020a}]{2020ApJ...897..123H}
{Hu} Y.,  {Lazarian} A.,   {Yuen} K.~H.,  2020a, \apj, \href
  {https://ui.adsabs.harvard.edu/abs/2020ApJ...897..123H} {897, 123}

\bibitem[\protect\citeauthoryear{{Hu}, {Lazarian}, {Li}, {Zhuravleva}  \&
  {Gendron-Marsolais}}{{Hu} et~al.}{2020b}]{2020ApJ...901..162H}
{Hu} Y.,  {Lazarian} A.,  {Li} Y.,  {Zhuravleva} I.,   {Gendron-Marsolais}
  M.-L.,  2020b, \mn@doi [\apj] {10.3847/1538-4357/abb1c3}, \href
  {https://ui.adsabs.harvard.edu/abs/2020ApJ...901..162H} {901, 162}

\bibitem[\protect\citeauthoryear{{Hu}, {Lazarian}  \& {Bialy}}{{Hu}
  et~al.}{2020c}]{2020ApJ...905..129H}
{Hu} Y.,  {Lazarian} A.,   {Bialy} S.,  2020c, \apj, \href
  {https://ui.adsabs.harvard.edu/abs/2020ApJ...905..129H} {905, 129}

\bibitem[\protect\citeauthoryear{{Hu}, {Xu}  \& {Lazarian}}{{Hu}
  et~al.}{2021a}]{2021ApJ...911...37H}
{Hu} Y.,  {Xu} S.,   {Lazarian} A.,  2021a, \apj, \href
  {https://ui.adsabs.harvard.edu/abs/2021ApJ...911...37H} {911, 37}

\bibitem[\protect\citeauthoryear{{Hu}, {Lazarian}  \& {Xu}}{{Hu}
  et~al.}{2021b}]{2021ApJ...915...67H}
{Hu} Y.,  {Lazarian} A.,   {Xu} S.,  2021b, \apj, \href
  {https://ui.adsabs.harvard.edu/abs/2021ApJ...915...67H} {915, 67}

\bibitem[\protect\citeauthoryear{{Hu}, {Lazarian}  \& {Xu}}{{Hu}
  et~al.}{2022}]{2021arXiv211115066H}
{Hu} Y.,  {Lazarian} A.,   {Xu} S.,  2022, \mn@doi [\mnras]
  {10.1093/mnras/stac319}, \href
  {https://ui.adsabs.harvard.edu/abs/2022MNRAS.tmp..341H} {512, 2111–2124}

\bibitem[\protect\citeauthoryear{{Kim} \& {Ryu}}{{Kim} \&
  {Ryu}}{2005}]{KimRyu2005}
{Kim} J.,  {Ryu} D.,  2005, \mn@doi [\apjl] {10.1086/491600}, \href
  {https://ui.adsabs.harvard.edu/abs/2005ApJ...630L..45K} {630, L45}

\bibitem[\protect\citeauthoryear{{Kitsionas} et~al.,}{{Kitsionas}
  et~al.}{2009}]{KitsionasEtAl2009}
{Kitsionas} S.,  et~al., 2009, \mn@doi [\aap] {10.1051/0004-6361/200811170},
  \href {https://ui.adsabs.harvard.edu/abs/2009A&A...508..541K} {508, 541}

\bibitem[\protect\citeauthoryear{{Klessen} \& {Hennebelle}}{{Klessen} \&
  {Hennebelle}}{2010}]{KlessenHennebelle2011}
{Klessen} R.~S.,  {Hennebelle} P.,  2010, \mn@doi [\aap]
  {10.1051/0004-6361/200913780}, \href
  {https://ui.adsabs.harvard.edu/abs/2010A&A...520A..17K} {520, A17}

\bibitem[\protect\citeauthoryear{{Klessen}, {Heitsch}  \& {Mac Low}}{{Klessen}
  et~al.}{2000}]{2000ApJ...535..887K}
{Klessen} R.~S.,  {Heitsch} F.,   {Mac Low} M.-M.,  2000, \mn@doi [\apj]
  {10.1086/308891}, \href
  {https://ui.adsabs.harvard.edu/abs/2000ApJ...535..887K} {535, 887}

\bibitem[\protect\citeauthoryear{{Kolmogorov}}{{Kolmogorov}}{1941}]{Kolmogorov1941c}
{Kolmogorov} A.,  1941, Akademiia Nauk SSSR Doklady, \href
  {https://ui.adsabs.harvard.edu/abs/1941DoSSR..30..301K} {30, 301}

\bibitem[\protect\citeauthoryear{Konstandin, Girichidis, Federrath  \&
  Klessen}{Konstandin et~al.}{2012}]{KonstandinEtAl2012}
Konstandin L.,  Girichidis P.,  Federrath C.,   Klessen R.~S.,  2012, The
  Astrophysical Journal, 761, 149

\bibitem[\protect\citeauthoryear{{Kowal} \& {Lazarian}}{{Kowal} \&
  {Lazarian}}{2010}]{2010ApJ...720..742K}
{Kowal} G.,  {Lazarian} A.,  2010, \mn@doi [\apj]
  {10.1088/0004-637X/720/1/742}, \href
  {https://ui.adsabs.harvard.edu/abs/2010ApJ...720..742K} {720, 742}

\bibitem[\protect\citeauthoryear{{Kritsuk}, {Norman}, {Padoan}  \&
  {Wagner}}{{Kritsuk} et~al.}{2007}]{KritsukEtAl2007}
{Kritsuk} A.~G.,  {Norman} M.~L.,  {Padoan} P.,   {Wagner} R.,  2007, \mn@doi
  [\apj] {10.1086/519443}, \href
  {https://ui.adsabs.harvard.edu/abs/2007ApJ...665..416K} {665, 416}

\bibitem[\protect\citeauthoryear{{Krumholz} \& {Burkhart}}{{Krumholz} \&
  {Burkhart}}{2016}]{KrumholzBurkhert2016}
{Krumholz} M.~R.,  {Burkhart} B.,  2016, \mn@doi [\mnras]
  {10.1093/mnras/stw434}, \href
  {https://ui.adsabs.harvard.edu/abs/2016MNRAS.458.1671K} {458, 1671}

\bibitem[\protect\citeauthoryear{{Krumholz} \& {McKee}}{{Krumholz} \&
  {McKee}}{2005}]{2005ApJ...630..250K}
{Krumholz} M.~R.,  {McKee} C.~F.,  2005, \mn@doi [\apj] {10.1086/431734}, \href
  {https://ui.adsabs.harvard.edu/abs/2005ApJ...630..250K} {630, 250}

\bibitem[\protect\citeauthoryear{{Larson}}{{Larson}}{1981}]{Larson1981}
{Larson} R.~B.,  1981, \mn@doi [\mnras] {10.1093/mnras/194.4.809}, \href
  {https://ui.adsabs.harvard.edu/abs/1981MNRAS.194..809L} {194, 809}

\bibitem[\protect\citeauthoryear{{Lazarian}}{{Lazarian}}{2004}]{2004JKAS...37..563L}
{Lazarian} A.,  2004, \mn@doi [Journal of Korean Astronomical Society]
  {10.5303/JKAS.2004.37.5.563}, \href
  {https://ui.adsabs.harvard.edu/abs/2004JKAS...37..563L} {37, 563}

\bibitem[\protect\citeauthoryear{{Lazarian}}{{Lazarian}}{2009}]{2009SSRv..143..357L}
{Lazarian} A.,  2009, \mn@doi [\ssr] {10.1007/s11214-008-9460-y}, \href
  {https://ui.adsabs.harvard.edu/abs/2009SSRv..143..357L} {143, 357}

\bibitem[\protect\citeauthoryear{{Lazarian} \& {Pogosyan}}{{Lazarian} \&
  {Pogosyan}}{2000}]{2000ApJ...537..720L}
{Lazarian} A.,  {Pogosyan} D.,  2000, \mn@doi [\apj] {10.1086/309040}, \href
  {https://ui.adsabs.harvard.edu/abs/2000ApJ...537..720L} {537, 720}

\bibitem[\protect\citeauthoryear{{Le Gouellec} et~al.,}{{Le Gouellec}
  et~al.}{2019}]{2019ApJ...885..106L}
{Le Gouellec} V. J.~M.,  et~al., 2019, \mn@doi [\apj]
  {10.3847/1538-4357/ab43c2}, \href
  {https://ui.adsabs.harvard.edu/abs/2019ApJ...885..106L} {885, 106}

\bibitem[\protect\citeauthoryear{{Lee} \& {Lee}}{{Lee} \&
  {Lee}}{2019}]{2019NatAs...3..154L}
{Lee} K.~H.,  {Lee} L.~C.,  2019, \mn@doi [Nature Astronomy]
  {10.1038/s41550-018-0650-6}, \href
  {https://ui.adsabs.harvard.edu/abs/2019NatAs...3..154L} {3, 154}

\bibitem[\protect\citeauthoryear{{Mac Low} \& {Klessen}}{{Mac Low} \&
  {Klessen}}{2004}]{MacLowKlessen2004}
{Mac Low} M.-M.,  {Klessen} R.~S.,  2004, \mn@doi [Reviews of Modern Physics]
  {10.1103/RevModPhys.76.125}, \href
  {https://ui.adsabs.harvard.edu/abs/2004RvMP...76..125M} {76, 125}

\bibitem[\protect\citeauthoryear{{MacNeice}, {Olson}, {Mobarry}, {de
  Fainchtein}  \& {Packer}}{{MacNeice} et~al.}{2000}]{2000CoPhC.126..330M}
{MacNeice} P.,  {Olson} K.~M.,  {Mobarry} C.,  {de Fainchtein} R.,   {Packer}
  C.,  2000, \mn@doi [Computer Physics Communications]
  {10.1016/S0010-4655(99)00501-9}, \href
  {https://ui.adsabs.harvard.edu/abs/2000CoPhC.126..330M} {126, 330}

\bibitem[\protect\citeauthoryear{{Mathew} \& {Federrath}}{{Mathew} \&
  {Federrath}}{2020}]{2020MNRAS.496.5201M}
{Mathew} S.~S.,  {Federrath} C.,  2020, \mn@doi [\mnras]
  {10.1093/mnras/staa1931}, \href
  {https://ui.adsabs.harvard.edu/abs/2020MNRAS.496.5201M} {496, 5201}

\bibitem[\protect\citeauthoryear{{Mathew} \& {Federrath}}{{Mathew} \&
  {Federrath}}{2021}]{2021MNRAS.507.2448M}
{Mathew} S.~S.,  {Federrath} C.,  2021, \mn@doi [\mnras]
  {10.1093/mnras/stab2338}, \href
  {https://ui.adsabs.harvard.edu/abs/2021MNRAS.507.2448M} {507, 2448}

\bibitem[\protect\citeauthoryear{{McKee} \& {Ostriker}}{{McKee} \&
  {Ostriker}}{2007}]{2007ARA&A..45..565M}
{McKee} C.~F.,  {Ostriker} E.~C.,  2007, \mn@doi [\araa]
  {10.1146/annurev.astro.45.051806.110602}, \href
  {https://ui.adsabs.harvard.edu/abs/2007ARA&A..45..565M} {45, 565}

\bibitem[\protect\citeauthoryear{{Nakamura} \& {Li}}{{Nakamura} \&
  {Li}}{2007}]{NakamuraLi2007}
{Nakamura} F.,  {Li} Z.-Y.,  2007, \mn@doi [\apj] {10.1086/517515}, \href
  {https://ui.adsabs.harvard.edu/abs/2007ApJ...662..395N} {662, 395}

\bibitem[\protect\citeauthoryear{{Ossenkopf} \& {Mac Low}}{{Ossenkopf} \& {Mac
  Low}}{2002}]{OssenkopfMacLow2002}
{Ossenkopf} V.,  {Mac Low} M.~M.,  2002, \mn@doi [\aap]
  {10.1051/0004-6361:20020629}, \href
  {https://ui.adsabs.harvard.edu/abs/2002A&A...390..307O} {390, 307}

\bibitem[\protect\citeauthoryear{{Padoan}}{{Padoan}}{1995}]{1995MNRAS.277..377P}
{Padoan} P.,  1995, \mn@doi [\mnras] {10.1093/mnras/277.2.377}, \href
  {https://ui.adsabs.harvard.edu/abs/1995MNRAS.277..377P} {277, 377}

\bibitem[\protect\citeauthoryear{{Padoan}, {Pan}, {Haugb{\o}lle}  \&
  {Nordlund}}{{Padoan} et~al.}{2016}]{2016ApJ...822...11P}
{Padoan} P.,  {Pan} L.,  {Haugb{\o}lle} T.,   {Nordlund} {\r{A}}.,  2016,
  \mn@doi [\apj] {10.3847/0004-637X/822/1/11}, \href
  {https://ui.adsabs.harvard.edu/abs/2016ApJ...822...11P} {822, 11}

\bibitem[\protect\citeauthoryear{{Pingel}, {Lee}, {Burkhart}  \&
  {Stanimirovi{\'c}}}{{Pingel} et~al.}{2018}]{2018ApJ...856..136P}
{Pingel} N.~M.,  {Lee} M.-Y.,  {Burkhart} B.,   {Stanimirovi{\'c}} S.,  2018,
  \mn@doi [\apj] {10.3847/1538-4357/aab34b}, \href
  {https://ui.adsabs.harvard.edu/abs/2018ApJ...856..136P} {856, 136}

\bibitem[\protect\citeauthoryear{{Qian}, {Li}, {Gao}, {Xu}  \& {Pan}}{{Qian}
  et~al.}{2018}]{2018ApJ...864..116Q}
{Qian} L.,  {Li} D.,  {Gao} Y.,  {Xu} H.,   {Pan} Z.,  2018, \mn@doi [\apj]
  {10.3847/1538-4357/aad780}, \href
  {https://ui.adsabs.harvard.edu/abs/2018ApJ...864..116Q} {864, 116}

\bibitem[\protect\citeauthoryear{{Ricker}}{{Ricker}}{2008}]{2008ApJS..176..293R}
{Ricker} P.~M.,  2008, \mn@doi [\apjs] {10.1086/526425}, \href
  {https://ui.adsabs.harvard.edu/abs/2008ApJS..176..293R} {176, 293}

\bibitem[\protect\citeauthoryear{{Roman-Duval}, {Federrath}, {Brunt}, {Heyer},
  {Jackson}  \& {Klessen}}{{Roman-Duval} et~al.}{2011}]{RomanDuvalEtAl2011}
{Roman-Duval} J.,  {Federrath} C.,  {Brunt} C.,  {Heyer} M.,  {Jackson} J.,
  {Klessen} R.~S.,  2011, \mn@doi [\apj] {10.1088/0004-637X/740/2/120}, \href
  {https://ui.adsabs.harvard.edu/abs/2011ApJ...740..120R} {740, 120}

\bibitem[\protect\citeauthoryear{{Schmidt}, {Federrath}, {Hupp}, {Kern}  \&
  {Niemeyer}}{{Schmidt} et~al.}{2009}]{SchmidtEtAl2009}
{Schmidt} W.,  {Federrath} C.,  {Hupp} M.,  {Kern} S.,   {Niemeyer} J.~C.,
  2009, \mn@doi [\aap] {10.1051/0004-6361:200809967}, \href
  {https://ui.adsabs.harvard.edu/abs/2009A&A...494..127S} {494, 127}

\bibitem[\protect\citeauthoryear{{Solomon}, {Rivolo}, {Barrett}  \&
  {Yahil}}{{Solomon} et~al.}{1987}]{SolomonEtAl1987}
{Solomon} P.~M.,  {Rivolo} A.~R.,  {Barrett} J.,   {Yahil} A.,  1987, \mn@doi
  [\apj] {10.1086/165493}, \href
  {https://ui.adsabs.harvard.edu/abs/1987ApJ...319..730S} {319, 730}

\bibitem[\protect\citeauthoryear{{Truelove}, {Klein}, {McKee}, {Holliman},
  {Howell}  \& {Greenough}}{{Truelove} et~al.}{1997}]{1997ApJ...489L.179T}
{Truelove} J.~K.,  {Klein} R.~I.,  {McKee} C.~F.,  {Holliman} John~H. I.,
  {Howell} L.~H.,   {Greenough} J.~A.,  1997, \mn@doi [\apjl] {10.1086/310975},
  \href {https://ui.adsabs.harvard.edu/abs/1997ApJ...489L.179T} {489, L179}

\bibitem[\protect\citeauthoryear{{Waagan}, {Federrath}  \&
  {Klingenberg}}{{Waagan} et~al.}{2011}]{WaaganEtAl2011}
{Waagan} K.,  {Federrath} C.,   {Klingenberg} C.,  2011, \mn@doi [Journal of
  Computational Physics] {10.1016/j.jcp.2011.01.026}, \href
  {https://ui.adsabs.harvard.edu/abs/2011JCoPh.230.3331W} {230, 3331}

\bibitem[\protect\citeauthoryear{{Xu} \& {Hu}}{{Xu} \&
  {Hu}}{2021}]{2021ApJ...910...88X}
{Xu} S.,  {Hu} Y.,  2021, \mn@doi [\apj] {10.3847/1538-4357/abe403}, \href
  {https://ui.adsabs.harvard.edu/abs/2021ApJ...910...88X} {910, 88}

\bibitem[\protect\citeauthoryear{{Xu} \& {Lazarian}}{{Xu} \&
  {Lazarian}}{2020}]{2020ApJ...890..157X}
{Xu} S.,  {Lazarian} A.,  2020, \mn@doi [\apj] {10.3847/1538-4357/ab6e63},
  \href {https://ui.adsabs.harvard.edu/abs/2020ApJ...890..157X} {890, 157}

\bibitem[\protect\citeauthoryear{{Xu} \& {Zhang}}{{Xu} \&
  {Zhang}}{2017}]{2017ApJ...835....2X}
{Xu} S.,  {Zhang} B.,  2017, \mn@doi [\apj] {10.3847/1538-4357/835/1/2}, \href
  {https://ui.adsabs.harvard.edu/abs/2017ApJ...835....2X} {835, 2}

\makeatother
\end{thebibliography}





\bsp	
\label{lastpage}
\end{document}